\documentclass[aps,prb,twocolumn]{revtex4}
\usepackage{mathrsfs, bm, amssymb}
\usepackage{graphicx,color}
\usepackage{amsmath}
\usepackage{fancyhdr}
\usepackage{graphicx}
\usepackage{epstopdf}
\usepackage{placeins}
\usepackage{float}
\usepackage{lmodern}
\usepackage[T1]{fontenc}
\usepackage{verbatim}
\usepackage{enumerate}

\def\onedot{$\mathsurround0pt\ldotp$}
\def\cdddot#1{
  \mathbin{\vcenter{\baselineskip.67ex
    \hbox{\onedot}\hbox{\onedot}\hbox{\onedot}%
  }}%
}
\def\fdot#1{
  \mathbin{\vcenter{\baselineskip.6ex
    \hbox{\onedot}\hbox{\onedot}\hbox{\onedot}\hbox{\onedot}%
  }}%
}
\usepackage{appendix}
\begin{document}
 \title[Nonlinear $XY$ and $p$-clock models] {Nonlinear $XY$ and
   $p$-clock models on sparse random graphs:
   mode-locking transition of localized waves} 

\author{Alessia
   Marruzzo$^{1,2}$ and Luca Leuzzi$^{2,1}$} 

\address{$^1$ Department
   of Physics, {\em La Sapienza} University, Piazzale Aldo Moro 2,
   Rome, Italy} 

\address{$^2$ Institute for Chemical-Physical
   Processes, IPCF-CNR, Rome Unit {\em Kerberos}, Piazzale Aldo Moro
   2, Rome, Italy}

\begin{abstract}
A statistical mechanic study of the $XY$ model with nonlinear
interaction is presented on bipartite sparse random graphs. The model
properties are compared to those of the $p$-clock model, in which the
planar continuous spins are discretized into $p$ values. We test the
goodness of the discrete approximation to the XY spins to be used in
numerical computations and simulations and its limits of convergence
in given, $p$-dependent, temperature regimes.  The models are applied
to describe the mode-locking transition of the phases of light-modes
in lasers at the critical lasing threshold. A frequency is assigned to
each variable node and function nodes implement a frequency matching
condition. A non-trivial unmagnetized phase-locking occurs at the
phase transition, where the frequency dependence of the phases turns
out to be linear in a broad range of frequencies, as in standard
mode-locking multimode laser at the optical power threshold.
\end{abstract} 
\maketitle

\section{Introduction}
\label{sec:intro}
 The $XY$ model with linearly interacting spins is well known in
 statistical mechanics displaying important physical insights and
 applications, starting from the Kosterlitz-Thouless transition in 2D,
 \cite{Kosterlitz72} and moving to, e.g., the transition of liquid
 helium to its super fluid state, \cite{Brezin82,Brezin10} the roughening
 transition of the interface of a crystal in equilibrium with its
 vapor~\cite{Cardy96} or synchronization problems related to the
 Kuramoto model. \cite{Kuramoto75,Acebron05, Gupta14} Furthermore, the
 $XY$ model with non-linear interaction terms has been used to
 investigate the topological properties of potential energy landscapes
 in configuration space. \cite{Angelani07} Our motivations to study
 non-linear $XY$ models are, though, to be found in optics, to
 describe, e.g., the non-linear interaction among electromagnetic
 modes in a laser cavity, \cite{Gordon02, Angelani07b, Antenucci14a}
 as well as the lasing transition in cavity-less amplifying resonating
 systems in random media known as {\em random lasers}.
 \cite{Angelani06,Leuzzi09, Conti11, Antenucci14b} Stimulated by this
 recent cross-fertilization of the fields of statistical mechanics and
 laser optics we are going to analyze a diluted $4$-body interacting
 $XY$-model on sparse random graphs including mode frequencies and
 gain profiles.

{\em Mode-}, or {\em phase-}locking~ \cite{LambBook} consists in the
amplification of very short pulses produced by the synchronization of
the phases of longitudinal axial modes in the cavity.  In the case of
{\em passive} mode-locking, yielding the shortest pulses,
synchronization is due to nonlinear mode-coupling. The most effective
known mechanism to induce nonlinearity is saturable absorption, that
is, the selective absorption of low intensity light and the
transmission of high intensity light leading, after many cavity
roundtrips, to a stationary train of ultra-short pulses. Such pulses
are composed by interacting modes of given, equispaced, frequencies
$\omega$ around a central frequency $\omega_0$.  In the typical case
of third order nonlinearity, \cite{LambBook,HausBook,Haus00} the modes
interact as quadruplets and must satisfy the Frequency Matching
Condition (FMC)
 \begin{equation}
\label{eq:frequency_matching}
 |\omega_j - \omega_k + \omega_l - \omega_m| \leq \gamma
 \end{equation}\noindent 
for each quadruplet composed by modes $(j,k,l,m)$,
 being $\gamma$ the line-width of the single mode. 
For such modes a constant {\em phase delay}
occurs, i.e., 
\begin{equation}
\phi(\omega) \simeq \phi(\omega_0) + \phi'\times
(\omega-\omega_0).
\label{eq:PL}
\end{equation} 
and the resulting electromagnetic signal is unchirped.  Mode phases
are, then, constrained as the relative frequencies by
Eq. (\ref{eq:frequency_matching}) and they are said to be {\em
  locked}.  If, as in standard laser cavities, resonaces are narrow
and evenly spaced, phases will be, thus, evenly spaced as well.  In
lasers with large gain band-width, the progressive depletion of low
intensity wings of the light pulse traveling through the cavity at
each roundtrip causes the amplification of very short pulses
composed by modes with locked phases.

When a laser operates in the multi-mode regime and reaches a
stationary state driven by the optical pumping, the interaction among
the modes can be described by the effective 4-mode interacting
Hamiltonian \cite{Gordon02,Gordon03,Antenucci14b}
 \begin{equation}
\label{eq:hamiltonian_modes}
 \mathcal{H} = -\Re \left[ \sum_{k} g_{k}a^*_k a_k + J
   \sum_{\{\omega_j,\omega_k,\omega_l,\omega_m\}}^{\rm FMC} a^*_j a_k a^*_l
   a_m \right]
 \end{equation}       
where $a_j \equiv A_j e^{i \phi_j}$ is the complex amplitude of the
light mode with eigenvector $\mathbf{E}_j(\mathbf{r})$, coefficient of
the following expansion for the electromagnetic field
\begin{equation}
 \mathbf{E}(\mathbf{r},t) = \sum_{j} a_{j}(t)
  \mathbf{E}_{j}(\mathbf{r}) e^{- i \omega_{j} t}  + \quad \mbox{c. c.}
\end{equation}

In the statistical mechanic approach, the total optical power pumped
into the system is required to be a constant of the problem, i.e., the
system is in a stationary, pumped driven regime effectively
representable as equilibrium phases in an adeguate ensemble.  The
total power is $\mathcal{E} = N \epsilon = \sum_k a_k a^*_k$. The
linear local coefficient $g_k$ in Eq. (\ref{eq:hamiltonian_modes}) is
the net gain profile and the non-linear coupling coefficient $J$
represents the self-amplitude modulation coefficient of the saturable
absorber responsible for the mode-locking regime. \cite{Haus00} It can
be expressed, as well, in terms of the spatial overlap of the
eigenvectors, \cite{Antenucci14a} i.e., given any four modes
$(j,k,l,m)$
\begin{equation}
J \propto \int d\bm r ~ \hat \chi^{(3)}(\bm r; \omega_j,
\omega_k,\omega_l, \omega_m) \fdot \mathbf{\bm E_j(\bm r) \bm E_k(\bm
  r) \bm E_l(\bm r) \bm E_m(\bm r)}
\label{eq:J_EEEE}
\end{equation}
where $\hat\chi^{(3)}$ is the nonlinear susceptibility tensor of the
optically active medium.

We will use the parameter $\beta$ as external driving force of the
transition.  In thermodynamic systems coupled to a thermal reservoir
at temperature $T$, $\beta=1/(k_B T)$ is simply the inverse
temperature.  In photonic systems it stand for an effective inverse
temperature related to both the real heat-bath temperature $T_{\rm
  bath}$ of the optically active medium and the optical power
$\epsilon$ pumped into the system as
\begin{equation}
\label{def:pumpingrate}
\beta J = \frac{\epsilon^2 J}{k_B T_{\rm bath}}\equiv {\cal P}^2
\end{equation}
where ${\cal P}$ is the so-called pumping rate.
\cite{Gordon02,Leuzzi09,Conti11,Antenucci14a}

The paper is organized as follows:
in Sec.  \ref{sec:model} we introduce the 4-XY and the 4-$p$-clock
models; in Sec.  \ref{sec:BP_CM} we recall the methods employed in the
analysis of the model and determine Belief Propagation and Cavity
equations for the specific models and in Secs.  \ref{sec:Bethe} and
\ref{sec:ER} we present the results on Bethe and on Erd\`os-R\'enyi
graphs. Eventually, in Sec.  \ref{sec:ML} we introduce a tree-like
mode-locking network and study the transition between the phase incoherent
regime and the coherent mode-locked regime typical of ultrafast
multimode lasers.


\section{$4$-XY model and $4-p$-clock model}
\label{sec:model}
The dynamic time-scales of magnitudes $\{A_j=|a_j|\}$ and 
phases $\{\phi_j=\arg(a_j)\}$
of the complex amplitudes are well separated. Since we are interested
in studying the phase-locking transition, we can consider observing the
system dynamics at a time-scale longer than the one of the phases but
sentively shorter than the one of the magnitudes, thus regarding the
amplitude magnitudes $A_k$ as constants. Within this {\em quenched
  amplitude} approximation, \cite{Angelani06,Leuzzi09} from
Eq. (\ref{eq:hamiltonian_modes}) we obtain 

\begin{equation}
\label{eq:hamiltonian}.
\mathcal{H} = - \sum_{jklm} J_{j k l m} \cos\left(\phi_j - \phi_k +
\phi_l - \phi_m \right)
\end{equation}
where we have rescaled $J A_jA_kA_lA_m \to J_{jklm}$. The sum
$\sum_{jklm}$ goes over the quadruplets for which the quenched
coefficients $J_{jklm}$ are different from zero, i.e., all quadruplets
whose electromagnetic fields overlap in space and whose frequencies
satify the FMC, Eq. (\ref{eq:frequency_matching}). The Hamiltonian
$\mathcal{H}$ is invariant under the $SO(2)$ group, i.e., rotations in
$2$ dimensions.  Imposing the further approximation that all
amplitudes are quenched {\em and} equal to each other, i.e., there is
intensity equipartition in every regime, one can define the
ferromagnetic nonlinear 4-XY model, $J_{jklm}=J$, $\forall (j,k,l,m)$,
whose behavior will be presented in this work on specific interaction
networks.

We will consider cases in which the number of interacting quadruplets
per mode does not grow with the size of the system. In terms of the
physical relationship between interaction coefficient and space
localization of modes, cf. Eq. (\ref{eq:J_EEEE}), this corresponds to
modes whose localization in space has an overall small volume but takes place
 in far apart, even disjoint, regions, yielding a dilute,
distance independent, interaction network.  These diluted model
instances will be represented as bypartite graphs.

Besides the XY-model, where spins are unitary vectors on a plane, $
\bm{\sigma} \equiv \left(\cos{\phi},\sin{\phi}\right)$, $ \phi\in
   [0,2\pi)$, we will consider a discretized version, where the phases
     $\phi$ can only take $p$ values, equispaced in radiants by $2
     \pi/p$:
\begin{equation}\label{eq:theta}
\phi_a = \frac{2 \pi}{p} a;  \qquad  a = 0,1,\dots,p-1
\end{equation}
We will use $p$ even, in order to be able to extend to the
antiferromagnetic and spin-glass cases, where the interactions among
spins can also be negative. Indeed, if $p$ is odd, it is not possible
to find a discretization of the $[0,2\pi)$ interval in such a way to
allow the four interacting spins to find the most energetically favorable
configurations for both $J>0$ and $J<0$. To better exemplify, if
$J<0$, a single  $(1,2,3,4)$ quadruplet contribution to the energy is
such that $\phi_1 + \phi_2 = \phi_3 + \phi_4 + \pi$. Discretizing
according to Eq. (\ref{eq:theta}) this implies $ a_1+a_2 =
p/2+a_3+a_4 $, that is effective only if $p$ is even.

The $p-$clock model can also be seen as a generalization of the Ising
model from $2$ to $p$ possible states for the local magnetization
$\sigma_i$: a spin varies over the $p$ roots of unity $e^{2 \pi
  a/p}$. The Hamiltonian Eq. (\ref{eq:hamiltonian}) is
invariant under the discrete symmetry group $Z_p$, consisting of
multiplying all the $\sigma_i$ by the same $p$th root. We know that in
the Ising case two phases can coexist when the symmetry $Z_2$ is
broken. In the $p>2$ case, there are $p$ phases that may coexist when
the symmetry is broken.  We will use the $p$-clock model as an
effectively tuned numerical representation for the $XY$ model.
Because the latter is a continuous model, we expect infinitesimal
fluctuations with infinitesimal energy cost to occur. These cannot be
present in a discrete model at low temperature: it is only in the $ p
\rightarrow \infty$ limit, thus, that we expect to recover all results
of the $XY$ model also in the $\beta \rightarrow \infty$ limit.  For
finite $\beta$, though, up to some extent the two representations
coincide.  In Secs. \ref{sec:Bethe} and \ref{sec:ER} we will
quantitatively determine such extent.  

Before presenting these
results, in the next section we are going to shortly recall the main tools
used, i.e., Belief Propagation, Cavity Method and Population Dynamical
Algorithm.  The paper is organized in such a way to let the reader
already familiar with these algorithms to skip Sec. \ref{sec:BP_CM}
and move  to Sec. \ref{sec:Bethe}.

\section{Belief propagation of the 4-XY model on factor graphs}
\label{sec:BP_CM}
We study the 4-XY model, Eq. (\ref{eq:hamiltonian}), on sparse random
graphs. In order to represent the 4-body interaction of phase
variables $\phi$ we, thus, resort to the factor graph representation
in terms of functional nodes of connectivity $k=4$ for the interacting
quadruplets and variable nodes of connectivity $c$ for mode phases
involved in $c$ quadruplets.  Let us label by $m=1,\ldots, M$ the
function nodes and by $\partial m$ the variable nodes connected to the
function node $m$. The phase $\phi_i$ is the value of the variable
node $i=1, \ldots, N$.

 A generic factor graph will be schematically
indicated by $G_N(k,M)$ where $N$ is the number of variable nodes, $M$
the number of function nodes (i.e., the number of interacting
$k$-uples), $Mk$ the number of edges connecting variable nodes to
function nodes and $\alpha = M/N= c/k$ is the
connectivity coefficient.
In general, we are  interested not only on  single
 instances, $G_N(k,M)$, but also on ensemble of factor
graphs.
We will focus on two general large groups:
random regular graphs, also known as Bethe lattices, and on Erd\`os
R\'enyi graphs.  Bethe graphs are
defined as follows: for each function node $m$ the $k-uple$ $\partial
m$ is taken uniformly at random from all the ${N \choose k}$ possibles
ones. In this case the fixed degree of connectivity $c$ of a variable
node is:
  \begin{equation}
  c = M \frac{{N \choose k-1}}{{N \choose k}} =
  \frac{Mk}{N-(k-1)}=\frac{Mk}{N}\left[1 +
    \mathcal{O}\left(\frac{k}{N}\right)\right]
  \end{equation}
in the diluted graph $k\ll N$.

 In Erd\`os R\'enyi graphs each $k$-uple is added to the factor graph
 independently, with probability $N \alpha/{N\choose k}$.  It can be
 proved \cite{Mezard09} that the total number of function nodes is a
 random variable with expected value $\langle M \rangle = N \alpha$
 while the degrees $c_i$ of the variable nodes are, in the large $N$
 limit, Poissonian independent identically distributed (iid) random
 variables with average $c=\langle c_i \rangle = \alpha k$.

 The factor graph representation for systems
described by Eq. (\ref{eq:hamiltonian}) yields the following joint
probability of a configuration of planar spins, i.e., phases 
${\bm\phi} =
\left(\phi_1,\phi_2,\ldots,\phi_N\right)$:
\begin{equation}\label{eq:joint}
 P\left(\bm \phi\right) = \frac{1}{Z} \prod_{m = 1}^{M} \psi_m
 \left({\phi}_{\partial m}\right)
\end{equation}
In order to find the equilibrium configurations of the system and
study the thermodynamic properties, we will use the Belief Propagation
(BP) method on factor graphs, $G_N(k,M)$, and the equivalent Cavity
Method (CM) for ensemble of random factor graphs. BP is an iterative
message-passing algorithm whose basic variables are messages
associated with directed edges.  For each edge ($i$,$m$) there exist
two messages $\nu^{(t)}_{i \rightarrow m}$ and $\hat{\nu}^{(t)}_{m
  \rightarrow i}$ that are updated iteratively in $t$ as
\begin{eqnarray}
\nonumber
 \hat{\nu}^{(t)}_{m \rightarrow i}\left(\phi\right) &=&  
 \frac{1}{z_{test}} \int_{0}^{2 \pi} \prod^{l=1,k-1}_{j_l \in \partial m \setminus i} 
d \phi_{j_l} \nu_{j_l \rightarrow m}^{(t-1)} \left(\phi_{j_l} \right) 
\\
\label{eq:uf}
&&\qquad\qquad\times 
 \psi_m \left(\phi_{j_1},\dots,\phi_{j_{k-1}},\phi \right) \\
 \label{eq:update_variab}
 \nu^{(t)}_{i \rightarrow m}\left(\phi\right) &=& 
 \frac{1}{z_{cav}} \prod_{n \in \partial i \setminus m} \hat{\nu}^{(t)}_{n \rightarrow i}\left(\phi\right)
\end{eqnarray}
where $\partial m = \{j_1,j_2,j_3,i\}$, $\partial m \setminus
i=\{j_1,j_2,j_3\}$, $\partial i$ indicates the neighbor function nodes
to variable node $i$ and $\partial i \setminus m$ are the function
nodes connected to $i$ but $m$.  $z_{test}$ and $z_{cav}$ are normalization 
factors. In the present XY model,
Eq. (\ref{eq:hamiltonian}), in which $k=4$, the Boltzmann weight
funtion is
\begin{equation}
\psi_m \left(\phi_{j_1},\phi_{j_2},\phi_{j_3},\phi\right) = e^{\beta J
  \cos \left(\phi_{j_1} - \phi_{j_2} + \phi_{j_3} - \phi \right)}
\label{eq:psi_m}
\end{equation}
 If the variable node $i$ is at
one end leaf of the graph, i.e., if $\partial i \setminus m$ is the
empty set, then it holds $\nu_{i \rightarrow
  m}\left(\phi\right)=1/(2\pi)$, the uniform distribution.\footnote{The
  effects of some external boundary can be described through the
  messages coming from the leafs. For example, if we want to consider
  a small external magnetic field, the $\nu_{i \rightarrow m}$ will
  depart from uniform on the external shell of nodes.}  
 BP equations are exact on
tree-like factor graphs.  When all message marginals, $\{\nu_{i
  \rightarrow m}$, $\hat{\nu}_{m \rightarrow i}\}$, are known, we can
evaluate the marginal probability distributions of the variable nodes:
 \begin{equation}
\mu_i(\phi_i) = \frac{1}{Z} \prod_{m \in \partial_i} \hat{\nu}_{m
  \rightarrow i} (\phi_i)
\end{equation}
The free energy of the system
reads\cite{Mezard09}
\begin{equation}\label{eq:final_free}
F = \sum_{m=1}^M F_m + \sum_{i=1}^N F_i - \sum_{im \in E} F_{im} 
\end{equation}
where $E$ indicates the set of all edges in the graph and 
\begin{eqnarray}
F_m & =& -\frac{1}{\beta}\log{\int_{0}^{2 \pi}\!\!\!\! \prod_{i \in \partial m}
 \! d \phi_i ~\nu_{i \rightarrow m}(\phi_i)~\psi_m\left(\phi_{\partial
    m} \right)} \\ 
F_i & =& -\frac{1}{\beta}\log{\int_{0}^{2 \pi} d
  \phi_i \prod_{m \in \partial i}\hat{\nu}_{m \rightarrow i}(\phi_i)}
\\ 
F_{im} & =& -\frac{1}{\beta}\log{\int_{0}^{2 \pi} d \phi_i ~
  \hat{\nu}_{m \rightarrow i}(\phi_i)~ \nu_{i \rightarrow m}(\phi_i)}
\end{eqnarray}

When we turn on ensembles of random factor graphs, the messages
$\nu_{i\rightarrow m}$ ($\hat{\nu}_{m\rightarrow i}$) become random
variables: the idea is then to use BP equations to characterize their
distributions in the large $N$ limit. Though BP equations are exact
only on tree-graphical models and sources of errors can come from the
existence of loops, they turn out to be a powerful tool on random
graphs, as well.  It is then useful to recall the results on the
probability of loops occurrence and their average length on Bethe and
Erd\`os R\'enyi graphs. It can be proved \cite{Mezard09} that, if
$\alpha k (k-1) < 1$, the fraction of nodes in finite size trees goes
to one as the total number of nodes $N$ goes to infinity: the
probability of having loops of any size goes to zero.  In the opposite
case, $\alpha k (k-1) > 1$, it appears in the graph what is known as
the ``giant component'': a connected part containing many loops.
Unlike the previous case, all the variable nodes belong almost surely
to this connected component. However, in the diluted case, loops have
infinite length and graphs look locally like trees.

 Being BP a local algorithm, one expects that, under the assumptions
 that correlations among variables go to zero as the distance between
 them diverges, a property termed {\em clustering}, \cite{MPV87} BP can be
 used to predict properties of the system in the thermodynamic limit.

 Then, for the case of random factor
 graphs, Eqs. (\ref{eq:uf}-\ref{eq:update_variab}) turn into
 equalities among the distributions $P(\nu)$, $Q(\hat{\nu})$ of the
 messages, i.e.,
\begin{eqnarray}
\label{eq:added_distribution_1}
\hat{\nu}(\phi) & \stackrel{d}{=}
&\frac{1}{z_{\rm test}} 
\int_{0}^{2\pi}
\prod_{l=1}^{k-1} d \phi_l ~\nu^{l}(\phi_l)
\\
&&
\nonumber\qquad\qquad\qquad\qquad\times 
\psi\left(\phi_{1},\dots,\phi_{k-1},\phi \right)
 \\
\label{eq:added_distribution_2}
 \nu\left(\phi\right) & \stackrel{d}{=} & \frac{1}{z_{\rm cav}}
 \prod_{m=1}^{c-1}\hat{\nu}^m\left(\phi\right)
\end{eqnarray}  
where $\nu^{l}$ and are $\hat{\nu}^{m}$ are i.i.d. marginal functions
and the connectivities $k$ and $c$ can, in principle, be random
variables.  The cavity method operates under the same assumptions we
have outlined above but suppose as well that
Eqs. (\ref{eq:added_distribution_1},\ref{eq:added_distribution_2})
have fixed-point solutions
$\{P^*(\nu),Q^*(\hat{\nu})\}$.\cite{Mezard01} Focusing on those
solutions, it evaluates recursively the partition functions by adding
one variable at a time. In fact, the term ``cavity'' comes from the
idea of creating a cavity around a variable by deleting one edge
coming from that variable. For example, consider a random graph $G$
where all edges coming from one constraint $m$ have been erased; call
$Z_{j \rightarrow m}(\phi_j)$ the partition function of one of the
$k$-tree graphs starting from one of the $j \in \partial m$ with
variable $j$ fixed to $\phi_j$; $Z_{j \rightarrow m}(\phi_j)$ can be
computed recursively:
 \begin{eqnarray}
 Z_{j \rightarrow m}(\phi_j) &=& \prod_{n \in \partial j \setminus m} 
 \Bigl[\prod_{i \in \partial n \setminus j} 
 \int_{0}^{2 \pi} d \phi_i
 ~\psi_n\left(\phi_{\partial n}\right)
\\
\nonumber
&&\qquad\qquad\qquad\times \prod_{i \in \partial n \setminus j}Z_{i \rightarrow n}(\phi_i)\Bigr]
 \end{eqnarray}
 BP equations (\ref{eq:uf},\ref{eq:update_variab}) are, then, obtained
 knowing the relation between BP messages and partition function:
 \begin{equation}\nonumber
 \nu_{j \rightarrow m} = \frac{Z_{j \rightarrow
     m}(\phi_j)}{\int_{0}^{2 \pi} d \phi_j Z_{j \rightarrow
     m}(\phi_j)}
 \end{equation}
 Once that the distributions of $\nu$ and $\hat{\nu}$ are known, the
 expected free-energy per variable $F/N$ can be computed taking the
 mean-value of equation (\ref{eq:final_free})
\begin{equation}\label{eq:final_tree_random}
 f = f_{\nu} + \frac{\overline{c}}{k}f_{\hat{\nu}} - \overline{c} f_{\nu~\hat{\nu}} 
\end{equation}
where 
\begin{eqnarray}
 f_{\nu} & =& -\frac{1}{\beta}\mathbb{E}_{c,\{\hat{\nu}\}} \left[
 \log{\int_{0}^{2 \pi} d \phi \prod_{m=1}^c \hat{\nu}^m(\phi)}\right]
\nonumber
\\
 f_{\hat{\nu}} & =& -\frac{1}{\beta}\mathbb{E}_{\{\nu\}} \left[
 \log{ \prod_{l=1}^k 
 \int_{0}^{2 \pi} \!\!d \phi_l ~\nu_l(\phi_l)\psi\left(\nu^1,\dots,\nu^k\right)}
 \right]
\nonumber
\\ 
 f_{\nu,\hat{\nu}} & =& -\frac{1}{\beta}\mathbb{E}_{\{\nu\},\{\hat{\nu}\}}\left[ \log{
 \int_{0}^{2 \pi} d \phi ~\nu(\phi)~\hat{\nu}(\phi)}\right] 
\nonumber \end{eqnarray}
and  $\mathbb{E}$ indicates expectation value with respect to the
variables in the subscript and $\overline{c}$ is the mean connectivity
of variable nodes.  Carrying out a functional derivative of
Eq. (\ref{eq:final_tree_random}), one can show that the stationary
points of the free-energy $f$ are in one-to-one correspondence with
solutions of BP equations.

The numerical method we use to solve Eqs.
(\ref{eq:added_distribution_1}, \ref{eq:added_distribution_2}) is
known in statistical physics as Population Dynamics Algorithm
(PDA). The idea is to approximate the distributions $P(\nu)$ and
$Q(\hat{\nu})$, through $N$ i.i.d. copies of $\nu$ and $\hat{\nu}$. We
call the sample $\{\nu_1,\dots,\nu_N\}$ (same for $\hat{\nu}$) a
population.  Starting from an initial distribution,
$\{\nu^0_1,\dots,\nu^0_N\}$, as the population evolves and its size is
large enough the distributions will converge to the fixed point
solution $\{P^*(\nu),Q^*(\hat{\nu})\}$. The convergence of the
algorithm is verified evaluating the statistical fluctuations of
intensive quantities.  Fluctuations of order $1/\sqrt{N}$ indicate the
convergence of the population to $\{P^*,Q^*\}$.\cite{Mezard09} Notice
that, for random regular graphs, since the connectivity is the same
for all nodes, if we take a functional identity initial distribution
$P(\nu) = \mathbb{I}\left(\nu - \nu_F\right)$, where $\nu_F$ is some
initial message, PDA is not necessary: we only have to consider the
updating of $\nu_F$.

In the next sections we will show the results obtained on
Bethe and ER graphs for different $p$ and $c$ values.  The results
presented have been obtained with population sizes up to $N = 6 \cdot
10^5$.

\section{XY- a $p$-clock models on random regular graphs}
\label{sec:Bethe}
In this section we will show the results obtained for the
ferromagnetic ($J=1$) 4-XY model on Bethe lattices: the degree of
variable nodes is fixed to $c$ while that of function nodes is $k=4$.
In order to numerically find the equilibrium distributions solving
Eqs. (\ref{eq:added_distribution_1}-\ref{eq:added_distribution_2}) for
the XY model, we resort to the discrete $p$-clock model,
cf. Eq. (\ref{eq:theta}).  Writing $\nu_a \equiv \nu(\phi_a)$, at
fixed $c$
Eqs. (\ref{eq:added_distribution_1}-\ref{eq:added_distribution_2})
become
\begin{eqnarray}\label{eq:uf_B_dis1}
  \hat{\nu}_a & \stackrel{d}{=}& \frac{1}{z_{\rm test}} \prod_{l=1}^3
  \left(\sum_{a_{l} = 0}^{p-1} \left(\nu^{l}\right)_{a_l}\right) 
\\
\nonumber
&&\qquad\qquad \qquad\qquad\times  e^{\beta J \cos{\frac{2 \pi}{p} \left(a_1-a_2+a_3-a\right)}}\\ 
\label{eq:uf_B_dis2}
\nu_a
  &  \stackrel{d}{=} & \frac{1}{z_{\rm cav}}
  \prod_{m=1}^{c-1}\left(\hat{\nu}^{m}\right)_{a}
\end{eqnarray}

In order to study possible fixed point solutions of
Eqs. (\ref{eq:uf_B_dis1},\ref{eq:uf_B_dis2}), it
is useful to introduce the Discrete Fourier Transform (DFT) of the message
$\nu$:
\begin{equation}\label{eq:dft_first}
 c_k = \sum_{a=0}^{p-1} \nu_a e^{\frac{- 2 \pi i k a}{p}}
\end{equation}
whose inverse transform is:
\begin{equation}\label{eq:dft_second}
 \nu_a = \frac{1}{p} \sum_{k=0}^{p-1} c_k e^{\frac{ 2 \pi i k a}{p}}   
\end{equation}
From Eq. (\ref{eq:dft_first}) we notice that
 $\nu_a$ is real, that is,
\begin{eqnarray}
\nonumber 
\sum_{k=0}^{p-1} c_k e^{\frac{ 2 \pi i k a}{p}}  
& =&\left(\sum_{k=0}^{p-1} c_k e^{\frac{2 \pi i k a}{p}}\right)^* 
 = \sum_{k=0}^{p-1} c^*_k e^{\frac{ 2 \pi i (p - k) a}{p}} 
\end{eqnarray}
 and $c_k = c^{*}_{p-k}$. In particular, $c_{{p/2}}$ is
 real. Furthermore, $c_0 = p/(2 \pi)$ and
 Eq. (\ref{eq:dft_second}) can be rewritten as
\begin{equation}\label{eq:dft_third}
 \nu_a = \frac{1}{ 2 \pi} \left(1 + \sum_{k = 1}^{p-1} \frac{2 \pi}{p}
 c_k e^{\frac{ 2 \pi i k a}{p}}\right)
\end{equation} 
Expressing the DFT of the cavity function in terms of magnitude and
phase, $c_k \equiv |c_k| e^{i \theta_k}$,
Eqs. (\ref{eq:uf_B_dis1}-
\ref{eq:uf_B_dis2}) becomes
\begin{eqnarray}
\label{eq:hat_p}
\hat{\nu}_a  &\stackrel{d}{=}&  \frac{1}{2\pi} 
 + \frac{1}{2 \pi p^3 I^p_0(\beta J)} 
\\
&&\qquad \times\Biggl[I^p_{{p/2}}(\beta J)
   \prod_{l=1}^{3}\left(c^{(l)}_{{p/2}} \right) \left(2 \pi\right)^3(-1)^a
  \nonumber  \\
 \nonumber &&\qquad \qquad  + \sum_{k=1}^{{p/2} - 1} I^p_k(\beta
   J) \left(\prod_{l=1}^{k-1}|c^{(l)}_k|\right) \left(2 \pi\right)^3 
\\
&&
\nonumber 
\qquad \qquad\times 2\cos
   \left(\theta^{(1)}_k - \theta^{(2)}_k + \theta^{(3)}_k + \frac{2 \pi a
     k}{p}\right) \Biggr]
  \end{eqnarray}
  where $I^p_k$ indicates the discrete approximation of the modified
  Bessel function of the first kind:
\begin{eqnarray}\label{eq:bessel_i}
  I^p_k(w) = \frac{1}{p} \sum_{a=0}^{p-1} e^{w \cos \left( \frac{2
     \pi a}{p} \right)} \cos \left(k \frac{2 \pi a}{p}\right)
 \end{eqnarray}
that, for $p\to\infty$, tends to the well-known
\begin{equation}
 I_k(w) = \frac{1}{2 \pi} \int_{0}^{2 \pi} e^{w
   \cos \phi} \cos \left(k \phi \right)
\nonumber
\end{equation}
Eq. (\ref{eq:hat_p}) is a distributional equality where
$c^{(1)},c^{(2)},c^{(3)}$ indicate the DFT of three i.i.d. $\nu$'s. It
can be observed that the trivial population distribution is $P(\nu) =
\mathbb{I}\left(\nu - \nu_{PM}\right)$, where $\left(\nu_{PM}\right)_a
=1/(2\pi) \quad \forall a$, i.e., when all $c^{(l)}_k = 0$, this is a
fixed point solution of
Eqs. (\ref{eq:uf_B_dis1}-\ref{eq:uf_B_dis2}) for
all values of $\beta J$. It is referred to as the
paramagnetic (PM) solution, invariant under $Z_p$ symmetry,
discretization of the SO(2) symmetry: there are no preferred
directions in the system and the spins are uniformly randomly
oriented.  We can notice that, as $p \rightarrow \infty$, we obtain
the correct, $SO(2)$ invariant, limit for the $XY$ PM solution
cf. Eqs. (\ref{eq:uf_B_dis1},\ref{eq:uf_B_dis2}).

 The fact that the uniform distribution is always a solution does not
 necessarily mean that the thermodynamic phase is always the 
 PM one. In given regions of the phase diagram,
 Eqs. (\ref{eq:uf_B_dis1},\ref{eq:uf_B_dis2})
 admit more than one fixed point solutions and the behavior of the
 model can be correctly described by a non-PM solution.  It is
 important to notice that any other solution for which at least one of
 the $c_k$ is different from zero is not invariant under
 $Z_p$. Therefore, if the system admits solutions other than the PM
 one, there will be spontaneous symmetry breaking.

In the case of a ferromagnetic (FM) solution, the system can align
itself among $p$ possible degenerate solutions, whose phases are
linked by the transformations of $Z_p$. Once the populations $P(\nu)$ and
$Q(\hat{\nu})$ are computed, we can evaluate the distribution of the
marginal probabilities of variable nodes:
\begin{equation}\label{eq:marginal}
 \mu(\phi) \stackrel{d}{=}
 \frac{1}{z_s}\prod_{l=1}^c\left[\hat{\nu}^l(\phi)\right]; \qquad
 z_{s} = \int_{0}^{2 \pi} d \phi
 \prod_{m=1}^c\left[\hat{\nu}^m(\phi)\right]
\end{equation}
  and, consequently, the magnetization, $m_x$ and $m_y$, and the
  free-energy, $f(\beta)$. In the continuous $p\to\infty$ limit we
  have for the  magnetization
\begin{eqnarray}\label{eq:magnetization}
 \langle m_x \rangle & =& \mathbb{E}_{\{\mu\}}\left(\int_{0}^{2 \pi}d\phi~
 \mu(\phi) \cos \phi\right)
 \\ 
\nonumber
\langle m_y \rangle & =&
 \mathbb{E}_{\{\mu\}}\left(\int_{0}^{2 \pi} d\phi~\mu(\phi) \sin \phi\right)
\end{eqnarray}
\noindent and for the free energy
\begin{equation}\label{eq:free_energy}
-\beta f\left(\beta\right) =\mathbb{E}_{\{\hat{\nu}\}}\log z_s 
+ \frac{c}{K} \mathbb{E}_{\{\nu\}}\log z_{c} - c 
 \mathbb{E}_{\{\nu,\hat{\nu}\}} \log z_{l}
\end{equation}
where 
\begin{eqnarray}
\nonumber
\label{eq:free_energy_sec}
 z_{c} & = & \int_{0}^{2 \pi}\left[\prod^4_{j=1}d\phi_{j}
   \nu^j(\phi_j) \right]e^{\beta
   J\cos\left(\phi_1-\phi_2+\phi_3-\phi_4 \right)} \\ 
\nonumber
z_{l} & = &
 \int_{0}^{2 \pi} d \phi~ \nu(\phi) ~\hat{\nu}(\phi)\\ 
\nonumber
z_{s} & = &
 \int_{0}^{2 \pi} d\phi~\left[\prod_{m=1}^c \hat{\nu}^m(\phi)\right]
\end{eqnarray}
For the $p-$clock free energy Eq. (\ref{eq:free_energy}) becomes:
\begin{eqnarray}\label{eq:free_energy_p}
  -\beta f\left(\beta\right) &=& \log \frac{2 \pi}{p} +
  \mathbb{E}_{\{\hat{\nu}\}} \log \sum_{a=0}^{p-1}
  \prod_{m=1}^c\left(\hat{\nu}^m_a\right) \\ 
\nonumber 
&& - c~
  \mathbb{E}_{\{\nu,\hat{\nu}\}} \log \sum_{a=0}^{p-1} \nu_a
  \hat{\nu}_a \\ 
&& \nonumber + \frac{c}{k} \mathbb{E}_{\{\nu\}}\log
  \prod^k_{j=1} \sum_{a_j = 0}^{p-1} \nu_{a_j}
\\
\nonumber
&&\qquad \times e^{\beta J \cos{\frac{2
        \pi}{p}\left(a_1-a_2+a_3-a_4\right)}}
\end{eqnarray}
In the PM solution ($m_x=m_y=0$) the
free-energy of the $p$-clock model is
\begin{eqnarray}\label{eq:free_energy_para}
f^p(\beta) & = & -\frac{1}{\beta}
  \left(\log 2 \pi + \frac{c}{k} \log I^p_0(\beta J) \right)
 \end{eqnarray}
where $I^p_0(w)$ is defined in Eq. (\ref{eq:bessel_i}). When a
solution other than the PM one appears, we may have $m_x$ or $m_y$ or
both different from zero: the total magnetization displays a preferred
direction and we have a FM solution.  The symmetry $Z_p$ is restored
if we notice that all the $p$ states can appear with the same
probability $1/p$ and we take the average over {\em pure states}:
\begin{equation}
\nonumber
m_{x,y} = \sum_{a = 0}^{p-1} \frac{1}{p} m^{p}_{x,y}
\end{equation}
 where $m^{p}_{x,y}$ are the magnetization values in state $p$. 

Eventually, the system at low temperature can be also found in a
{\em phase-locked} (PL) phase, where $m_{x,y}=0$ but phases are nevertheless
locked into a non-trivial relation among them, i.e.,  
not only Eq. (\ref{eq:magnetization}) is zero but also 
\begin{equation}
\int d\phi~ \mu(\phi) \cos\phi= \int d\phi~ \mu(\phi) \sin\phi=0, \quad \forall
\mu(\phi) \, .
\nonumber
\end{equation}
When this occurs, the order parameter to spot such a phase is
\begin{equation}
\label{eq:q}
r=2\mathbb{E}_{\{\mu\}}\int d\phi ~\mu(\phi)~ \cos^2\phi - 1
\end{equation}
This is trivially equally to $0$ in the paramagnetic phase but it
acquires a different value $r\in[-1,1]$ when the system is in the PL
phase.

\subsection{$p$-clock convergence to XY}
 
 Considering Eq. (\ref{eq:hat_p}) we will derive the main features of
 the solutions as a function of the number of clock-states $p$. We
 can, thus, check what is the minimum number of values of the $XY$
 angle to obtain an effective description of the model with continuous
 $XY$ spins.  The 4-$XY$ PM/FM phase transition, unlike the case with
 only two body interaction terms ($k=2$) \cite{Lupo14} turns out to be
 first order, discontinuous in internal energy and in order
 parameters.  In general, the number $p$ guaranteeing convergence
 between $p$-clock and $XY$-models will depend on the temperature
 range.  In particular, we will compare (i) spinodal points, (ii)
 paramagnetic free energies { and (iii) ferromagnetic free energies}
 to establish convergence of the two models. (i) Indicating as $\beta_s$
 the inverse temperature of the FM spinodal, as $p$ increases it holds
 $\beta_s^{p+2} \ge \beta_s^p$. This derives from the fact that,
 cf. Eq. (\ref{eq:bessel_i}),
\begin{equation}
\label{eq:beta_s}
R_{01}^p(x)\equiv \frac{I^{p}_1(x)}{I^{p}_0(x)}-
\frac{I^{p+2}_1(x)}{I^{p+2}_0(x)} \geq 0 .
\end{equation} 

\begin{figure}
 \includegraphics[width=.99\columnwidth]{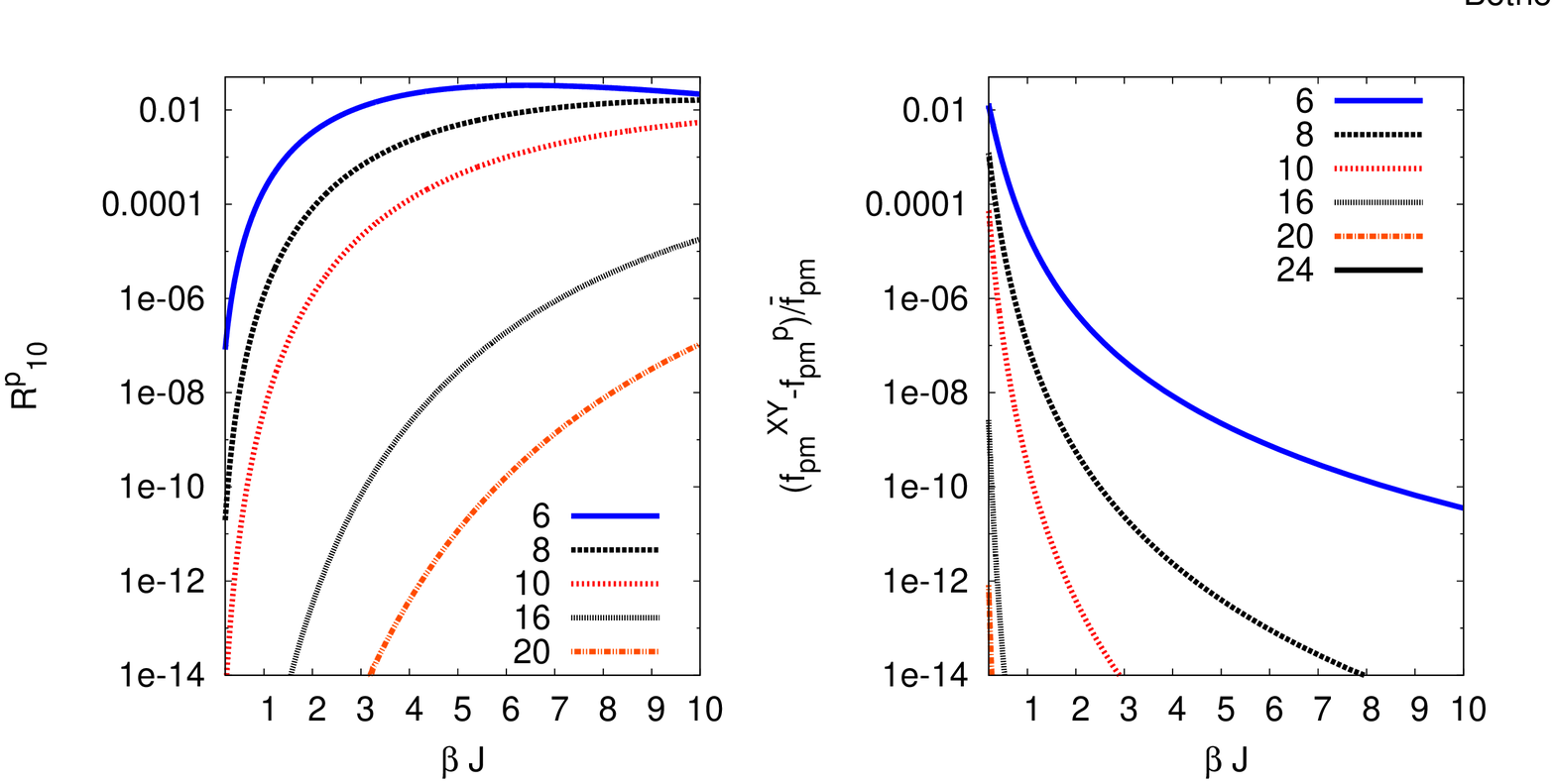}
 \caption{ (Left) $R^{p}_{01}$ as a function of $\beta J$ for
   $c=6$. We can see that as $p$ increases the convergence to the XY
   model holds up to larger and larger $\beta$ values. For $p=20$
   ($p+2=22$), the difference among the two is smaller than double
   precision up to values of $\beta J \simeq 3$. (Right) Convergence
   of the $p$-clock paramagnetic free energy to the $XY$ paramagnetic
   free energy for $c=6$.  The denominator is $\bar f_{\rm pm
   }(\beta)\equiv (f_{\rm pm}^{p}+f_{\rm pm}^{XY})/2$. The relative
   difference between the two decreases with $\beta J$.}
 \label{fig:ratio_i}
\end{figure}

\begin{figure}
 \includegraphics[width=.99\columnwidth]{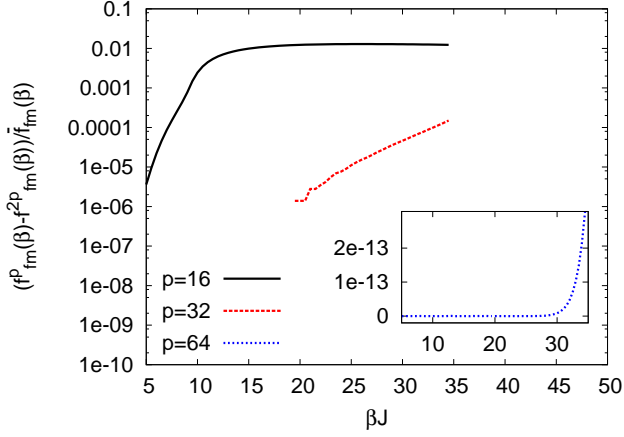}
 \caption{ Relative free energy difference of the FM phase for
   different $p$, $2p$ couples of $p$-clock models. The denominator is
   $\bar f_{\rm fm}(\beta)\equiv (f_{\rm fm}^{(p)}+f_{\rm
     fm}^{(p+2)})/2$. Already for $p=16$ and $p=32$ the relative
   difference saturates at $10^{-2}$ for $\beta J >10$. As shown in the
   inset practically no difference can be appreciated in double
   precision between $p=64$ and $p=128$ up to $\beta J\simeq 30$. }
 \label{fig:f_fm_conv}
\end{figure}

The behavior of Eq. (\ref{eq:beta_s}) is plotted in the left panel of
Fig. \ref{fig:ratio_i}. (ii) The paramagnetic free-energy can be
computed analytically, both for $p$-clock and the continuous
$XY$-model. We can, therefore, evaluate the number of spin states,
$p$, needed to converge to the $XY$ model in the desired temperature
interval also from the PM free enegy difference, cf. right panel of
Fig. (\ref{fig:ratio_i}). (iii) In Fig. \ref{fig:f_fm_conv} the
numerical comparison of the ferromagnetic free energy is shown between
$p$-clock models with, respectively $p$ and $2p$ states.  As it
becomes clear in the inset, already for $p=64$ no difference can be
further appreciated for very high $\beta$ values, much larger than the
critical $\beta_c$, as it will soon be shown.  We also stress that at
very low temperature a direct comparison with the XY-model free energy
cannot be performed, because the latter continuos model has an
ill-defined entropy at $T=0$ and its free energy is, thus, defined
expect for a constant. Comparison with the $XY$ model, thus, implies
the necessity of introducing a ($p$-dependent) constant.

\begin{figure}
 \includegraphics[width=.99\columnwidth]{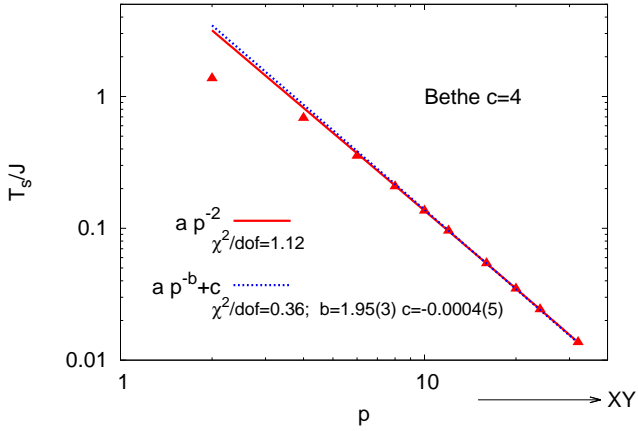}
 \caption{ Spinodal temperature, $T_s$, as a function of $p$ with its
   best fits for vanishing $T_s$ in the $p\to\infty$ limit for fixed
   connectivity $c=4$. For continuous $XY$ spins, when $T>0$, the only
   fixed point solution of $BP$ equations is the paramagnetic
   solution. The fact that we obtain a ferromagnetic solution is an
   artifact induced by $p< \infty$. }
 \label{fig:fit_exp_beta_c_n}
\end{figure}

\begin{figure}
 \includegraphics[width=.99\columnwidth]{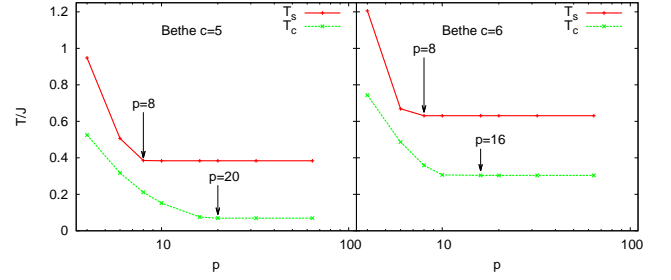}
 \caption{Spinodal and critical temperatures $T_s$ and $T_c$ vs $p$ in
   the Bethe lattice, for $c=5$ (left) and $c=6$ (right): the values
   of $p$ for which convergence to the $XY$ limit is
   attained are marked by arrows.}
 \label{fig:beta_c_n_56}
\end{figure}

In Figs. \ref{fig:fit_exp_beta_c_n}, \ref{fig:beta_c_n_56} we report
the results obtained for spinodal and critical point as a function of
the number of states $p$ for different values of the connectivity
$c=4,5,6$. At the critical inverse temperature $\beta_c=1/T_c$ the PM
solution becomes metastable.  As we can see from
Fig. \ref{fig:fit_exp_beta_c_n}, the lower critical connectivity for
the $XY$ model is $c_{\rm low}=5$: FM solutions for $c=4$ are an
artifact of taking $\phi$ as a discrete variable.  In
Fig. \ref{fig:beta_c_n_56} we show the convergence to the XY limit in
$p$ for $c=5,6$. The convergence is faster for the spinodal point but
not much slower for the critical point: $p\simeq 20$-clock spin is
already a rather good approximation of the planar continuous spin for
what concerns the analysis of the critical behavior.

\subsection{Critical behavior of the 4-XY model }

\begin{figure}[b!]
 \includegraphics[width=.99\columnwidth]{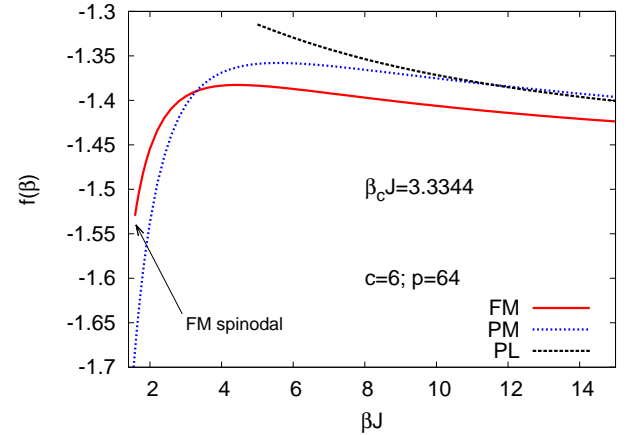}
 \caption{Free energy, $f(\beta)$ vs.  $\beta J$ for $c=6$ and
   $p=64$. {The full line} refers to the ferromagnetic fixed
   point solution, found considering as initial conditions the effect
   of a strong external magnetic field. { The dotted line}  refers to the
   paramagnetic solution. Unlike in the $k=2$
   case, the paramagnetic solution is stable at every temperature.
The dashed line represents the metastable phase-locked solution.}
 \label{fig:f_beta}
\end{figure}

\begin{figure}[b!]
 \includegraphics[width=.99\columnwidth]{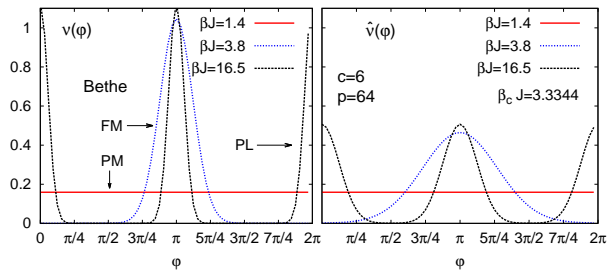}
 \caption{$\nu_a$ and $\hat{\nu}_a$ for $p=64$ and three different
   phases at three values of $\beta J$: PM at $\beta J= 1.4<\beta_cJ$,
   FM at $\beta J=3.8>\beta_c J$ and PL at $\beta J=16.5$. }
 \label{fig:messages}
\end{figure}

We thus study the properties of the XY model across the critical point
using a $p=64$ clock model.  In Fig. \ref{fig:f_beta} we display the
free energy for $c=6$ as a function of $\beta J$ for the three fixed
point solutions of BP equations (\ref{eq:dft_third}-\ref{eq:hat_p}):
PM, FM and PL phases. The FM solution is selected by tuning the
initial conditions assigning higher probability to a given $\phi$
value. The PL solution is obtained at high enough $\beta$ when initial
$\nu(\phi)$ are given with two peaks at opposite angles.
In Fig. \ref{fig:messages} we report the resulting marginal cavity
distributions for the phase values, $\nu(\phi)$ and $\hat \nu(\phi)$.

In the PL phase, though at each local instance $m_{xy}=0$, the
parameter $r$ defined in Eq. (\ref{eq:q}) is not.  Its free energy
behavior is shown in Fig. \ref{fig:f_beta} as dashed line. It can be
observed that the PL phase is always metastable with respect to the FM
phase, though, for higher $\beta$ its free energy becomes lower than
the PM free energy.  Because of the observed numerical fragility of
such solution with respect to the PM and the FM phases, it is hard to
discriminate its spinodal point.  With the computation performed so
far the PL phase appears to occur for $\beta J\gtrsim 4.5$.

\section{XY- and $p$-clock models on Erd\`os R\'enyi factor graphs}
\label{sec:ER}

If the degrees of variable nodes are i.i.d. random variables, the
local environment is not the same everywhere in the graph.  In the
Erd\`os R\'enyi case BP equations are \emph{distributional equations}
as in Eqs. (\ref{eq:uf_B_dis1}-\ref{eq:uf_B_dis2}) where the number of
neighbors to a variable node are extracted by means of a Poissonian
distribution of average $c$.

In this section we show the results obtained by applying the PDA on
the ordered $p$-clock model on ER graphs and look for asymptotic
solutions as $p \rightarrow \infty$.  The results presented have been
obtained with a population size up to $N = 6 \cdot 10^5$.  The code used to
numerically determine $\nu, \hat{\nu}$ stationary populations for
large $p$ is a parallel code running on GPU's. This sensitively speeds
up the population update of $\hat{\nu}$ (requiring $N p^k$ operations)
with respect to a serial, CPU running, code.

\begin{figure}[t!]
\includegraphics[width=.99\columnwidth]{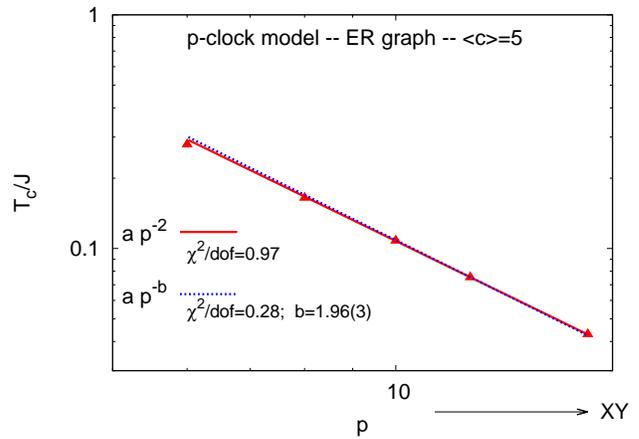}
\caption{Spinodal point values $T_s/J$ vs. $p$ on Erd\`os R\'enyi
  factor graphs with mean connectivity $\langle c \rangle = 5$. The
  interpolations displayed are both consistence with the absence of a
  magnetized phase in the XY, for  $T>0$.}
\label{fig:beta_s_5}
\end{figure}

\begin{figure}[t!]
\includegraphics[width=.99\columnwidth]{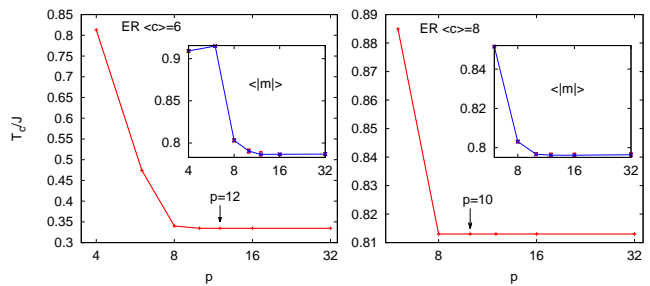}
\caption{Critical point as a function of integer $p$ on Erd\`os
  R\'enyi factor graphs with mean connectivity $\langle c \rangle =
  6$ (left) and $8$ (right). 
  In the insets we show the absolute values of the magnetization as a
  function of $p$. }
\label{fig:beta_s_6}
\end{figure}

In Fig. \ref{fig:beta_s_5} we show the values obtained for $T_s/J$
when the mean connectivity of variable nodes is $\langle c \rangle =
5$.  We can see that in this case the only solution in the $p
\rightarrow \infty$ limit is the PM solution, whereas other solutions
with $\langle m^2 \rangle \neq 0$ are artifacts of $p < \infty$. 

In Fig. \ref{fig:beta_s_6} we report the results obtained when
$\langle c \rangle =6$ and $8$: as for regular random graphs the
convergence to the $XY$ model is rather fast (see also the inset for
the absolute value of the magnetization).

We observe that $c_{\rm low}=6$ for the ER graph is larger than the
corresponding value $c_{\rm low}=5$ in the Bethe lattice. The presence
of many nodes with connectivity $\langle c\rangle -1$ or lower, when
$\langle c\rangle =5$, apparently leads to a zero transition
temperature in the ER graph. We notice, however, that in the linear
case ($k=2$) the trend is the opposite: for Bethe lattices the minimal
connectivity for a non-trivial critical behavior is $c_{\rm low}=3$,
for Erd\`os R\'enyi graphs is $c_{\rm low} = 2$, as reported in
App. \ref{appendix:C}. More details on the linear case can be found in 
Ref. [\onlinecite{Lupo14}].


\section{Mode-locking on  random graphs}
\label{sec:ML}
As mentioned in the introduction, the nonlinear $XY$ model can be used
to describe the phase dynamics of interacting electromagnetic modes in
lasers. Previous mean-field studies on fully connected models assume
{\em narrow-band} for the spectrum, \cite{Angelani06, Angelani07,
  Leuzzi09,Conti11} that is, all modes practically have the same
frequency and, in this way, the frequencies do not play any role in
the system behavior.  This is the case for the systems analyzed in
Secs. \ref{sec:Bethe} and \ref{sec:ER}. In this section, exploiting the
diluted nature of the graphs, we deepen such description and allow for
the existence of finite-band spectra and gain frequency profiles.

\begin{figure}[t!]
\centering
\includegraphics[width=.99\columnwidth]{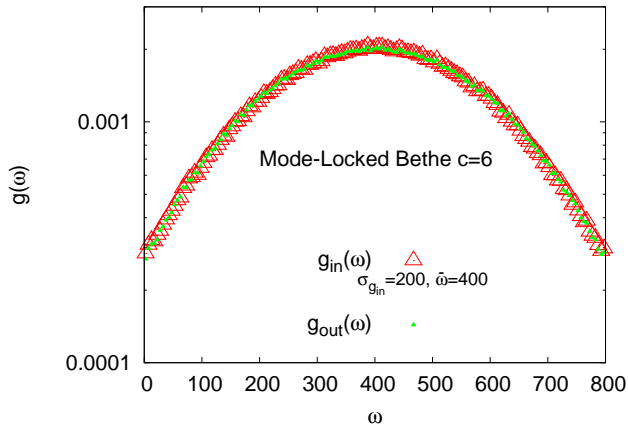}
\caption{Empty triangles refer to the distribution of frequencies
  assigned to $2/3$ of the variable nodes, according to a Gaussian
  gain profile, $g_{in}(\omega)$ of mean $\bar\omega=400$ and variance
  $\sigma_{g_{\rm in}}=200$. Filled-in triangles refer to the
  distribution $g_{\rm out}(\omega)$ we obtain once the FMC is
  imposed: $g_{\rm out}(\omega)$ coincides with $g_{in}(\omega)$ on
  the whole domain. }
\label{fig:gain_profile}
\end{figure}

Tree-like factor graphs can be built where each variable node,
representing a light mode, has a quenched frequency associated to its
dynamic phase.  The frequencies are distributed among modes according
to, e.g., a Gaussian or a parabolic distribution proportional to the
{\em optical gain} $g(\omega)$ for the system resonances.  The graph
is, then, constructed starting from the root in such a way that the
FMC Eq. (\ref{eq:frequency_matching}) is satisfied for each
interacting quadruplet.  Else said, a function node $m$ {\em is} a FMC
for the $\{\partial m\}$ modes connected to it.  As an example, in
Fig. \ref{fig:gain_profile}, we show a possible frequency distribution
for a tree-like factor graph in which the connectivity of the variable
nodes is fixed to $c=6$.  The empty (large) triangles refer to the
gain profile, $g_{in}(\omega)$, according to which the frequencies are
assigned to free variable nodes ($2/3$ of the total). The
remaining $1/3$ of the node frequencies are assigned according to the
FMC. Note that applying the FMC one can obtain three possible
independent combinations for the fourth frequency. From Fig.
\ref{fig:gain_profile} we can see that the frequency distribution of
all frequencies, $g(\omega)$, evaluated once that the FMC has been
imposed for the all quadruplets, is compatible with the starting one,
$g_{in}(\omega)$. This result shows that, considering a generic
Gaussian gain profile, sparse factor graphs, in which $\langle c
\rangle = \mathcal{O}(1)$, can yield a meaningful realistic
description of non-linearly interacting modes whose frequencies
satisfy the FMC.

\subsection{Phases and phase-locking}

\begin{figure}[t!]
\includegraphics[width=.99\columnwidth]{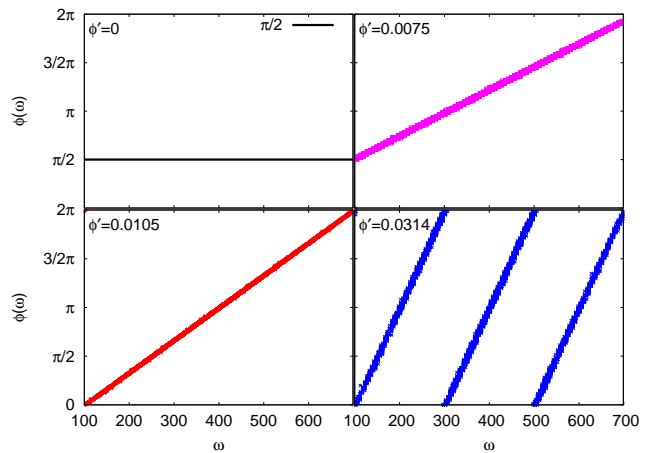}
\caption{Phases vs. frequencies in phase-locked phases on ML-Bethe
  lattice with $c=6$, $N_{\rm shell}=5$ shells, and a total number of
  inner nodes (excluding leaves) $N_{\rm bulk}=4339$. The number of
  clock tics is $p=120$. The number of frequencies is $N_\omega=88$ or
  $120$. The pumping rate squared is ${\cal P}^2=\beta J=7$.  }
\label{fig:PL}
\end{figure}

Once graphs with fixed connectivity and frequency matching function
nodes are introduced we can study the critical behavior considering
$\beta J$ as a pumping rate squared ${\cal P}^2$, 
cf. Eq. (\ref{def:pumpingrate}), in the context of lasing systems.
We will term these
graphs ``Mode-Locking Bethe'' (ML-Bethe) lattices.  As a result of BP,
above a certain threshold of ${\cal P}$ mode phases turn
out to show a peculiar behavior in the frequencies:  $\phi(\omega)$ 
 coincides with the
linear law of Eq. (\ref{eq:PL}), as shown in Fig. \ref{fig:PL} for
different linear coefficients $\phi'$.  Though, generically, the
magnetizations are $m_{xy}=0$, the phases are, nevertheless, found to
be locked.  This is the typical behavior established at the lasing
threshold by nonlinearity in multimode lasers.  In the above mentioned
construction of the ML-Bethe lattice, frequencies are assigned to
modes with a probability proportional to the gain profile.

 We take into acccount two qualitatively different cases.  First we
 consider the case where only equispaced frequencies are eligible:
 this is a proxy for the so-called {\em comb } distribution
 \cite{Bellini00,Udem02} in which many resonances occurs with a
 line-width much smaller than the fixed resonance interspacing.
Furthermore, we investigate the opposite  extreme, the  {\em
  continuous} case, in which each mode frequency is extracted continuously from
the whole gain band with no further constraint on their values, other than
FMC.

\begin{figure}[t!]
\includegraphics[width=.99\columnwidth]{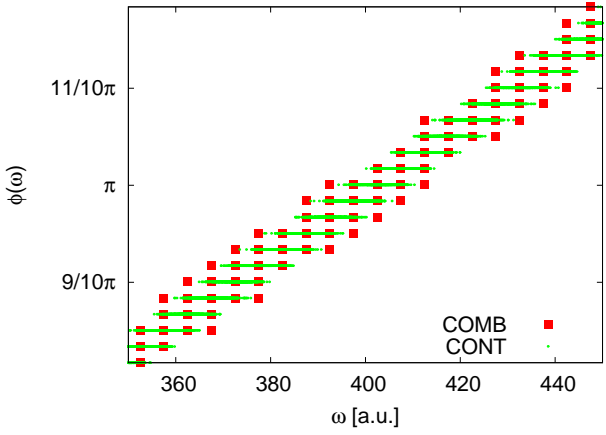}
\caption{Detail of the behavior of the phases vs. frequencies
  extracted by means of the distribution of
  Fig. \ref{fig:gain_profile} both as continuous and as comb-like
  equispaced at $\beta J = 7$, for $c=6$, $N_{\rm shell}=4339$, $N_{\rm
    bulk}=65089$, $p=120$, for $120$ continuous (light grey/green points) and
  comb (dark grey/red squares) frequencies.}
\label{fig:PL_cont_vs_comb}
\end{figure} 
In Fig. \ref{fig:PL} we show different realizations of such
phase-locking, all of them with different {\em phase delay}
$\phi'$. They are frequency independent and do not depend on
frequencies being equispaced or continuously distributed.  This
amounts to say that phase delay disperion is zero.  Each locking is
obtained by means of different boundary conditions at the external
shell.  The case $\phi'=0$ is also achieved, that is the ferromagnetic
phase: all modes are locked at the same phase.  In term of
thermodynamics all realizations of phase-locking, including the
ferromagnetic one, display comparable free energies, all of them
definitely different from the free energy of the coexisting PM phase.

Altough phase-locking, cf.  Eq. (\ref{eq:PL}), occurs in both the comb
and the continuous frequency distributions, as shown in
Fig. \ref{fig:PL_cont_vs_comb} there is a difference in the range of
values that frequencies can take at each (discrete) value of the
phases.  We anticipate that only in the case of comb-like distributions of
gain resonances mode-locking allows to realize {\em ultrashort
  pulses}.

\begin{figure}[t!]
\includegraphics[width=.99\columnwidth]{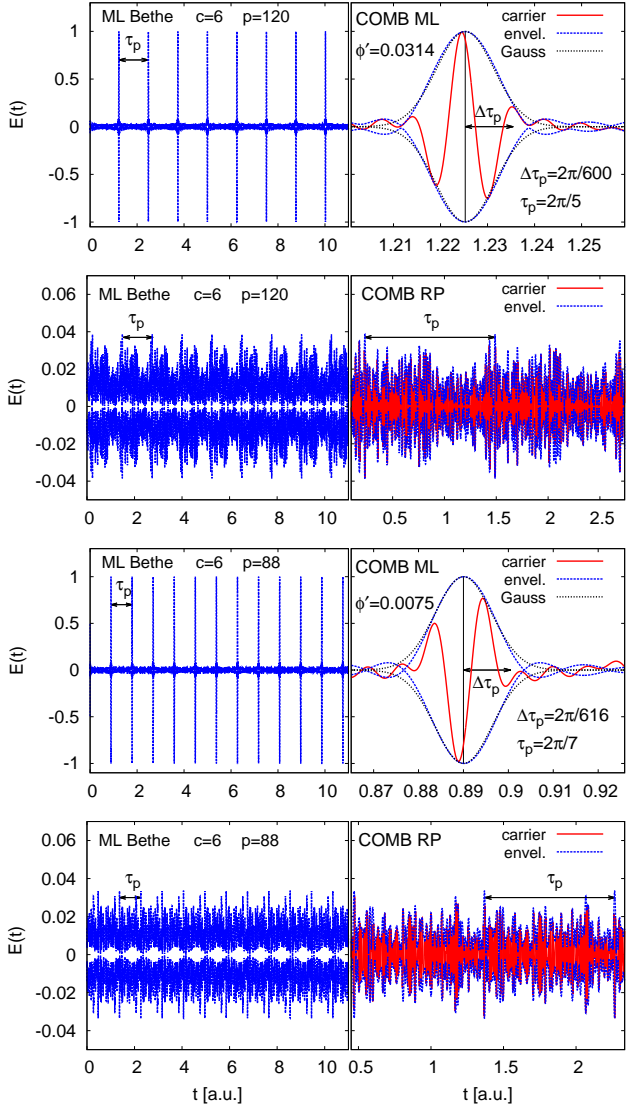}
\caption{The laser pulse $E(t)$ generated in the lasing phase in a ML
  Bethe lattice with comb-like frequency distribution. Two different
  realizations of the phase-locking are reported, with delay
  $\phi'=0.0314$ (top four panels) and $\phi'=0.0075$ (bottom four
  panels).  In the left panels several periodc pulses are shown, with
  a period $\tau_p=2\pi/\Delta \omega$, with $\Delta \omega=5$ in the
  top case and $\Delta\omega=7$ in the bottom case.  In the right
  panels the details of the single pulse are given, where both carrier
  and envelope are plotted. In the ML pulsed phase (first and third
  right panels) the pulse half-width is $\Delta \tau_p=2
  \pi/(N\Delta\omega)$, where $N=120$ for $\phi'=0.0314$ and $N=88$ for
  $\phi'=0.0075$. We also plot the behavior of the amplitude $\pm
  |E(t)|$ expected for Gaussian gain profiles, i.e., the square root
  of (\ref{eq:em_Gauss}), displaying a rather good coincidence. }
\label{fig:pulses}
\end{figure}

We, eventually, come to the analysis of the electromagnetic signal for a
wave system with $N=N_{\rm bulk}$ modes and $N_\omega$ frequencies:
\begin{eqnarray}
E(t) &=&\sum_{k=1}^N A_k e^{\imath (\omega_k t + \phi_k)}
\nonumber
\\
&=& 
e^{\imath (\omega_0 t +\phi_0) }
\sum_{k=1}^N A_k e^{\imath (\Delta \omega_k t + \Delta \phi_k)}
\end{eqnarray} 
where the sinusoidal carrier wave frequency $\omega_0$ is the central
frequency of the spectrum (of the order of $10^{15}$ rad $\cdot$
s$^{-1}$), $\phi_0=\phi(\omega_0)$ and $\Delta\omega_k$'s are of the
order of radio frequencies (ca. $10^9$ rad $\cdot$ s$^{-1}$). In the
ML regime, where, cf. Eq. (\ref{eq:PL}), $\Delta \phi_k=\phi' \Delta
\omega_k$ the time dependent overall amplitude can be written as
\begin{eqnarray}
A(t)&\equiv&\sum_{k=1}^N A_k e^{\imath (\Delta \omega_k t + \Delta \phi_k)}
\label{eq:a_t}
\\
&=&\sum_{k=1}^N A_k e^{\imath \Delta \omega_k (t + \phi')}=A(t+\phi')
\nonumber
\end{eqnarray}
The term phase (or group) delay for $\phi'$ comes from the fact that
it corresponds to a shift in time in the $E(t)$ carrier peak with
respect to the $|E(t+\phi')|=|A(t+\phi')|$ {\em envelope} maximum.
If, furthermore, $N_\omega$ comb distributed resonances are considered
with interspacing $\Delta \omega$, we can write
\begin{eqnarray}
\label{eq:nl}
&&e^{\imath \Delta\omega_k t} = n_l e^{\imath l \Delta\omega t}; \\
\nonumber &&\quad k=1,\ldots, N;
\quad l=-N_\omega/2,\ldots,N_\omega/2-1
\end{eqnarray}
where $n_l$ is the number of modes at frequency $l\Delta\omega$.  This
is the case for ultra-short ML lasers for which very short and very
intense periodic pulses occur, as shown for $N_{\rm bulk}=4339$ modes
in the first and third left panels in Fig. \ref{fig:pulses} for
$\phi'=0.0314$ ($N_\omega=120$) and $\phi'=0.0075$ ($N_\omega=88$),
respectively.

Since we are working in the quenched amplitude approximation with
intensity equipartition each mode has magnitude $A_k=1$.  However, we
are using  diluted interaction networks and, consequently, the same frequency
can be taken by modes
localized in different spatial regions, whose number we denote by $n_l$
in Eq. (\ref{eq:nl}).  Therefore,
\begin{eqnarray}
\label{eq:em_nl}
E(t)=
e^{\imath (\omega_0 t +\phi_0) }
\sum_{l=-N_\omega/2}^{N_\omega/2-1} n_l ~e^{\imath l \Delta \omega (\phi'+t)}
\end{eqnarray} 
and, from the point of view of the Fourier decomposition of the e.m. signal,
$n_l$ plays the role of the amplitude of the modes at frequency $l$.

A detail of the pulses is shown in the first and third right panels of
Fig.  \ref{fig:pulses}.  The linear behaviors shown in
Fig. \ref{fig:PL}, alike to Eq. (\ref{eq:PL}), implies that the signal
is {\em unchirped}. Else said, the phase delay displays no dispersion
and the frequency of oscillation of the carrier remains the same for
all pulses, as can be observed  in the right
panels of Fig.  \ref{fig:pulses}.  The period of the pulses is
$\tau_p=2\pi/\Delta \omega$, where $\Delta \omega=5$ for phase delay
$\phi'=0.0314$ and $\Delta \omega=7$ for $\phi'=0.0075$.  The pulse
duration is expressed in terms of its Full Width Half Maximum $\Delta
\tau_p$, also equal to the time it takes for the e.m. field amplitude
$A(t)$,  to decrease to zero from its maximum.

In ML ultrafast lasers, if the gain has a Gaussian profile in the equidistant
frequencies, and ,consequently,  $n_l$ is so distributed, cf. 
Eq. (\ref{eq:em_nl}), the signal amplitude squared is expected to behave like
\begin{equation}
|E(t)|^2=|E(t_{\rm max})|^2\exp\left\{-\left(2\frac{t-t_{\rm max}}{\Delta\tau_p}
\right)^2\ln 2\right\}
\label{eq:em_Gauss}
\end{equation}
in the limit of very many frequencies ($\Delta \omega\to
0$).\cite{Svelto98} In Fig. \ref{fig:pulses}, first and third right
panels, this behavior is plotted as ``Gaussian''.  Above the noise
level it appears to coincide very well with the envelope obtained by
Fourier Transform of the output of BP equations on ML Bethe lattices.

 In the second ($\phi'=0.0314$) 
and fourth ($\phi'=0.0075$) rows of Fig. \ref{fig:pulses} we show $E(t)$
 in the low pumping paramagnetic phases, where modes display random
 phases (RP).  The periodicity induced by the comb-like distribution
 appears also here, though the electromagnetic field is purely noisy,
 without any pulse.

\begin{figure}
\centering
\includegraphics[width=.99\columnwidth]{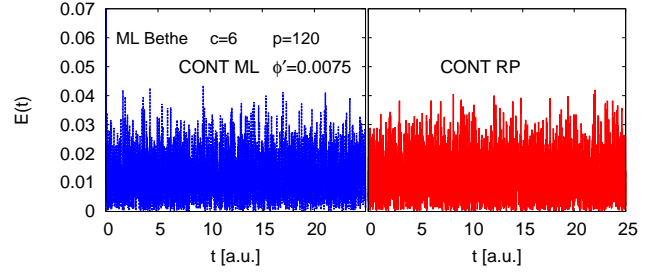}
\caption{The amplitude of the e.m. field, $|E(t)|$,
 cf. Eq. (\ref{eq:a_t}), is plotted for a ML
  Bethe lattice with continuously frequency distribution.
In the first row we display the $\phi'=0.0314$ case in the high pumping 
mode-locked regime (left) and in the low pumping
random phase regime (right).
In the second row $\phi'=0.0075$. }
\label{fig:nopulses}
\end{figure}

When frequencies are taken in a continuous way the coherent
phase-locked phase turns out to display much less intense coherent
signal, with no pulses, as shown in Fig. \ref{fig:nopulses} where
$|E(t)|$ is shown both in the ML and in the random phase
(paramagnetic) regimes, with no apparent difference in the time domain
between coherent and incoherent light.

\section{Conclusions}
In the present work we have undergone the investigation of the $XY$
model with non-linear, $4$-body, interaction and of its discrete
approximant, the so-called $p$-clock model, on random graphs.  Cavity
equations have been derived and solved for the Bethe lattice and the
Erd{\'o}s-R{\'e}nyi graph, carrying out a thourough analysis of the
critical behavior in temperature at varying connectivity values. Three
phases are found for these models. At high $T$ the systems are in a
paramagnetic phase. At low $T$ the dominat thermodynamic phase is
 ferromagnetic, that
is, a $SU(2)$ continuous symmetry breaking occurs in the XY model and
a $Z_p$ discrete symmetry breaking occurs in the $p$-clock
model. Else, a low temperature metastable {\em phase-locking } phase
can be reached, in which the magnetization is zero but the phases,
though all different, are nevertheless correlated to each other. 
An accurate study of the convergence of the $p$-clock model to the 
continous model is performed and presented.

 The models introduced can be applied to laser optics, where the $XY$
 or $p$-clock spins play the role of light mode phases. In this
 photonic framework the inverse temperature $\beta$ is proportional to
 the square of the rate of population inversion, the so-called {\em
   pumping rate}, driving the lasing transition from the incoherent
 light regime.  The first result is that a mode-locking Bethe lattice
 can be consistently built in which, besides the phase, also a
 frequency is associated to each variable node and each function node
 acts as a {\em frequency matching condition} among four frequencies,
 cf. Eq. \ref{eq:frequency_matching}. The latter is a common kind of
 non-linear interaction occurring in standard ultra-fast multimode
 lasers. As $\beta$ increases the system is found to undergo a
 mode-locking transition: phases at nearby frequencies are locked to
 take fixed amount and a linear $\phi(\omega)$ relationship like
 Eq. (\ref{eq:PL}) is established at the critical point. In the case
 of evenly distributed mode frequencies this leads to a pulsed laser,
 i.e., a laser whose electromagnetic field oscillations are
 characterized by a train of very short and very intense pulses. We
 have been comparing the results obtained in this case to the laser
 signal for multimode frequencies randomly taken in a continuous
 dominion, as well as to the incoherent signal below the lasing
 threshold.  The model presented, thus, provides an analytical and
 phenomenologically accurate description of multimode lasers at the
 level of the {\em single} pulse, that can be chosen arbitrarily
 shorter than the period between two pulses when the frequencies are
 evenly spaced as, e.g., in standard Fabry-Perot cavities. Such a
 limit is not achievable experimentally because the typical response
 time of conventional photodetectors is of the order of $1$ ns,
 whereas the duration of pulses in ultra-fast mode-locking solid-state
 or semiconductor lasers ranges from the order of the picosecond to
 the order of the femtosecond.  

Eventually, laser emission is also investigated in the opposite
extreme, where frequencies can take any value according to a given
gain profile, not only evenly spaced values. These systems undergo
phase-locking, because of the frequency matching condition, but prove
a far less intense signal, more akin to the signal of early
continuous-wave pumped solid-state lasers. \cite{Nelson62} Such
frequency limit distribution is, in principle, compatible with the
random topology of light localizations on sparsely connected
interaction networks that can represent a salient feature of more
complex laser systems called random lasers.\cite{Lawandy94, Cao99,
  Cao05, Wiersma08} In these systems, indeed, where also the magnitude
and even the sign of the mode coupling can be disordered, the
pumping rate threshold values are known to be higher and the signal
intensity is found to be sensitively smaller than in standard {\em
  ordered} multimode lasers.

 \section*{Acknowledgements} The research
leading to these results has received funding from the Italian
Ministry of Education, University and Research under the Basic
Research Investigation Fund (FIRB/2008) program/CINECA grant code
RBFR08M3P4 and under the PRIN2010 program, grant code 2010HXAW77-008
and from the People Programme (Marie Curie Actions) of the European
Union's Seventh Framework Programme FP7/2007-2013/ under REA grant
agreement n. 290038, NETADIS project.


\appendix

\section{XY model with linear interaction on sparse graphs}
\label{appendix:C}

Let us consider the two point correlation function for the $XY$ model
with pairwise interaction, $k=2$:
\begin{equation}
 \mathcal{H} = -\sum_{(i,j)} J_{ij} \cos{\left(\phi_i - \phi_j\right)}
\end{equation}
Taking two variable nodes, $i$ and $j$, we indicate by $U_{ij}$ the
shortest path that goes from $i$ to $j$, by $F_R$ the subset of
function nodes (now simple links) in $U_{ij}$ and by $V_R$ the subset
of variable nodes in $U_{ij}$ including $i$ and $j$. Then let
$\partial R$ be the subset of function nodes that are not in $U_{ij}$
but are adjacent to the variable nodes in $V_R$: $\forall m \in
\partial R$, $\exists'$ $l \in \partial m \bigcap V_R$, which will be
called $l(m)$.  Then, we have that the joint probability distribution
of all variables in $R$ is:
\begin{equation}
\label{eq:joint_correlation}
 \mu\left(\underline{\phi_R}\right) = \frac{1}{Z_R}\prod_{m \in F_R} \psi_m\left(\phi_{\partial m}\right) \prod_{m \in \partial R} %
 \hat{\nu}_{m \rightarrow l(m)}\left(\phi_{l(m)}\right)  
\end{equation}
We, then, denote by $r$ the distance between the two initial spins $i$
and $j$. The distance $r$ is, in fact, the number of links in $F_R$,
each one with its marginal $\hat{\nu}$. Recalling
Eq. \ref{eq:uf} we obtain that, in the paramagnetic phase
(where $\hat{\nu}(\phi) = \frac{1}{2 \pi}$, $\forall \phi\in[0,2\pi)$), the 
two-spin joint probability
distribution function is:
\begin{eqnarray}
\label{eq:joint_2xy}
 \mu(\phi_i,\phi_j) &=& \frac{1}{(2 \pi)^2}+\frac{1}{(\pi~I_0(\beta J))^r} 
\\
\nonumber
&&\quad \times \sum_{n=1}^{\infty} \left(I_n(\beta J)\right)^r \cos(n (\phi_i 
- \phi_j)) 
\end{eqnarray}
Consequently, it is 
\begin{equation*}
\langle \cos(\phi_i) \cos(\phi_j) \rangle = \langle \sin(\phi_i) \sin(\phi_j) \rangle = \frac{1}{2}
\left(\frac{I_1(\beta J)}{I_0{\beta J}}\right)^r
\end{equation*}
and
\begin{equation}
 C(r)\equiv\langle \cos(\phi_i-\phi_{i+r}) \rangle = \left(\frac{I_1(\beta
   J)}{I_0(\beta J)}\right)^r
\end{equation}
The susceptibility can be written as
\begin{equation}
 \chi = \frac{1}{N}\sum_{r=0}^{\infty} \sum_{(i,j=i+r)\in \mathcal{G}} \langle \sigma_i \sigma_{j} \rangle
 = \sum_{r=0}^{\infty} \mathcal{N}(r)
 \left(\frac{I_1(\beta J)}{I_0(\beta J)}\right)^r
\end{equation}
where $(i,i+r)$ indicates all the links in the graph
between two variable nodes at distance $r$ and $\mathcal{N}(r)$ is the
expected number of variables nodes $j$ at a distance $r$ from a
uniformly random node $i$. For large $r$ on a Bethe graph
$\mathcal{N}(r) = (c-1)^r$ and $\chi < \infty$ for 
\begin{equation}
\left(\frac{I_1(\beta J)}{I_0(\beta J)}\right)(c-1)<1
\end{equation}
As \begin{equation}
\label{eq:stability_k_2}
\left(\frac{I_1(\beta_c J)}{I_0(\beta_c J)}\right)(c-1)=1
\end{equation}
$\chi=\infty$, the paramagnetic solution becomes unstable and the
value of the critical temperature $T_c = \frac{1}{\beta_c}$ is
determined.

For the case of Erd\`os R\'enyi graphs, we obtain for large $r$ $
\mathcal{N}(r) = \langle c \rangle = c $ where we used the property of
the Poissonian distribution $P_c(k)$, with $k=c-1$:
\begin{eqnarray}
\sum_{c=2}^\infty (c-1) P_c(c-1) &=& \sum_{c=2}^\infty e^{-c}
\frac{c^{c-1}}{(c-1)!}(c-1) 
\\
&=& c~ e^{-c} \sum_{c=2}^\infty
\frac{c^{c-2}}{(c-2)!} = c
\nonumber
\end{eqnarray}
Then, in this case Eq. \ref{eq:stability_k_2} becomes:
\begin{equation}
\label{eq:stability_k_2_poisson}
 \left(\frac{I_1(\beta_c J)}{I_0(\beta_c J)}\right)(c)=1
\end{equation}
We stress that both critical conditions Eqs. \ref{eq:stability_k_2}
and \ref{eq:stability_k_2_poisson} can be obtained by expanding Eq.
\ref{eq:uf} around the paramagnetic solution
\cite{Skantzos05,Lupo14}. In the ER case we see that the presence of
nodes with connectivity larger than $c$ has the effect of lowering
$\beta_c$, i.e. increasing $T_c$.



\end{document}